\def\-{~-~}
\def\+{~+~}
\def\={~=~}

\def\beq{\begin{equation}}
\def\eeq{\end{equation}}
\def\beqn{\begin{eqnarray}}
\def\eeqn{\end{eqnarray}}
\def\bb{\begin{eqnarray*}}
\def\ee{\end{eqnarray*}}
\def\blet{\begin{mathletters}}
\def\elet{\end{mathletters}}
\newcommand{\calle}[1]{(\ref{#1})}
\documentstyle[preprint,revtex,eqsecnum]{aps}
\begin{document}
\draft
\preprint{CLNS-93/1206}
\medskip
\begin{title}
\bf{ A Comment on the Propagator \\
of the Radial Oscillator}
\end{title}

\author{\bf{Costas John Efthimiou}$^\dagger$}

\begin{instit}
{Newman Laboratory of Nuclear Studies}\\
Cornell University\\
{Ithaca, NY 14850 USA}
\end{instit}

\begin{abstract}
Using  a hybrid approach, based on the  recursion relations
for shape invariant potentials developed by Das and Huang and
a time-dependent tranformation
of variables, we derive
the propagator for a radial oscillator. Although this is not a new result,
 we explicitly
show that time-dependent tranformations are very beneficial even within
the context of time-independent Hamiltonians in quantum mechanics.

\end{abstract}

\vfill
\hrule
$\dagger$ {costas@beauty.tn.cornell.edu}\\
\newpage
After the introduction of Feynman path integrals in quantum mechanics
\cite{Feynman} a lot of work has ben undertaken to evaluate them for
many  systems. Today we know the  path integrals for all soluble
potentials of quantum
mechanics. (For a recent review see reference \cite{Grosche}).
 In fact, more than one methods have been proposed for most of
these potentials in order to calculate the corresponding path integrals.
 The purpose of this note is to present one more method for the potential
of the radial oscillator
\beq
\label{radialoscillator}
  V(x)\={1\over2}\,m\,\omega^2\,x^2\+{\hbar^2\over 2m}\,{n(n+1)\over x^2}~,
  ~~~~~~n=1,2,3,\dots~.
\eeq
(Be aware of the terminology: we work in one space dimension).
This potential is known to be soluble by shape invariance \cite{Gendenshtein}
with superpotential:
$$
        W(x)\=\sqrt{m}\omega\,x\+{\hbar\over\sqrt{m}}\,{n\over x}~.
$$
The $n$-th hamiltonial of the corresponding series of shape invariant
potentials is a simple harmonic oscillator
$$
       H^{(n)}\=-{\hbar^2\over 2m}\, {\partial^2\over\partial x^2}\+
  {1\over2}\,m\,\omega^2\,x^2\+{\hbar\omega\,\left(2n-{1\over2}\right)}~,
$$
for which the propagator is well-known:
\beqn
        K^{(n)}(x,x';t)\= \sqrt{m\omega\over2\pi i\hbar{\rm sin}\omega t}~
e^{-{i\omega t}\left(2n-{1\over2}\right)}
 \Bigg\{\, && {\rm exp}\left\lbrack-{im\omega \over2\hbar}
  \left( (x^2+x'^2){\rm cot}\omega t-{2xx'\over{\rm sin}\omega t}
  \right)   \right\rbrack
\nonumber\\
  -&&{\rm exp}\left\lbrack-{im\omega \over2\hbar}
  \left( (x^2+x'^2){\rm cot}\omega t+{2xx'\over{\rm sin}\omega t}
  \right)   \right\rbrack\,\Bigg\}~.
\nonumber
\eeqn
Notice that because the radial oscillator potential is singular,
 we must restrict
the motion on the half line $0\le x < +\infty$; this induces to the propagator
the additional term due to the reflection at the origin \cite{Pauli}.

One way then to derive the corresponding propagator of the initial potential
\calle{radialoscillator} is  to use the recursion
relations of Das and Huang \cite{DasHuang}
\beqn
  \left\lbrack i\hbar\,{\partial\over\partial t}-2s\omega\hbar\, \right\rbrack
      K^{(s)}(x,x';t)\={1\over2}&&
  \left\lbrack{\hbar\over\sqrt{m}}\,
   {\partial\over\partial x}-W(x)\right\rbrack\,\nonumber\\
  &&\left\lbrack{\hbar\over\sqrt{m}}\,
   {\partial\over\partial x'}-W(x')\right\rbrack\,
      K^{(s+1)}(x,x';t)~.
\label{recursion}
\eeqn
Notice that in order to derive the $s$-th propagator fron the $(s+1)$-th one,
we must perform an integration over  the time  variable;
although this integration is straightforward,
because of the trigonometric functions involved, it  is a little cumbersome.
In \cite{DasHuang},\cite{Khare}, using the  of recursion relation
\calle{recursion} for $s=0$,
 the exact propagator for the radial oscillator in the case $n=1$ has been
found although it  is not written  in the usual standard form (in terms of a
Bessel function) but it is expressed in an equivalent form.

In order to derive the propagator for the radial oscillator for a general $n$,
we propose the following alternative way.
Using the phase transformation
\beq
\label{phasetransformation}
     \Psi(x,t)\={1\over \sqrt{{\rm cos}\omega t}}\,
         {\rm exp}\left(-{im\omega\over2\hbar}\,{\rm tan}\omega t
\,x^2\right)\,
         \Phi(x,t)~,
\eeq
and the change of variables
\beq
\label{changevariables}
     \tau\={{\rm tan}\omega t \over\omega }~,~~~~~y\={x\over{\rm cos}\omega
t}~,
\eeq
the Schrodinger equation of the radial oscillator
\beq
\label{xSch}
   i\hbar{\partial\Psi\over\partial t}\=
   -{\hbar^2\over 2m}\, {\partial^2\Psi\over\partial x^2}
   +\left({1\over 2}\, m\, \omega^2\, x^2 +{\hbar^2\over 2m}\,{n(n+1)\over x^2}
    \right)\,\Psi~,
\eeq
simplifies to
\beq
\label{ySch}
   i\hbar{\partial\Phi\over\partial\tau}\=
   -{\hbar^2\over 2m}\, {\partial^2\Phi\over\partial y^2}
   +{\hbar^2\over 2m}\,{n(n+1)\over y^2}\,\Phi~.
\eeq
For this transformed problem, the frequency of the oscillator is zero and
the $n$-th hamiltonian of the series of shape invariant potentials
is the one of the free particle:
$$
    H^{(n)}\=
   -{\hbar^2\over 2m}\, {\partial^2\Phi\over\partial y^2}~.
$$
The propagator for this last hamiltonian is just the free propagator, including
an additional term due to reflection at the origin:
$$
     K(y,y';\tau)\=\sqrt{m\over 2\pi\hbar i\tau}\,
     \left\lbrack\, e^{im(y-y')^2\over2\hbar\tau}
     \- e^{im(y+y')^2\over2\hbar\tau}\,\right\rbrack~.
$$
Starting with the above propagator,
the integrations over the time variable
in the recursion relations of Das and Huang are much easier to be carried out.
They  have already been done in \cite{DasHuang},\cite{Khare}.
The result is then given by:
\beqn
     K(y,y';\tau)\=\sqrt{m\over 2\pi\hbar i\tau}\,
     \Bigg\lbrack\, e^{im(y-y')^2\over2\hbar\tau}&\,&\sum_{p=0}^n\,
     {(n+p)!\over p!(n-p)!}\, \left({-i\hbar\tau\over2myy'}\right)^p
    \nonumber\\
     &+&(-1)^{n+1}\, e^{im(y+y')^2\over2\hbar\tau}\,\sum_{p=0}^n\,
     {(n+p)!\over p!(n-p)!}\, \left({i\hbar\tau\over2myy'}\right)^p
     \Bigg\rbrack~~~
\eeqn
Seperating the odd from the even terms in the above summations,
we can rewrite this result in the form
\beqn
     K(y,y';\tau)\=\sqrt{{m\over 2\pi\hbar i\tau}}\,(-2i^{1-n})
          e^{{im\over 2\hbar\tau}(y^2+y'^2)}\,
     \Bigg\{\, \sum_{p=0}^{\lbrack {n\over2} \rbrack}\,
     {(-1)^p\,(n+2p)!\over (2p)!(n-2p)!(2z)^{2p}}\,{\rm sin}\left(
    z-{n\pi\over 2}\right)
    \nonumber\\
     + \sum_{p=0}^{\lbrack {n-1\over2} \rbrack}\,
     {(-1)^p\,(n+2p+1)!\over (2p+1)!(n-2p-1)!(2z)^{2p+1}}\,{\rm cos}\left(
    z-{n\pi\over 2}\right)\Bigg\}~,~~~~
\label{prop1}
\eeqn
where $[x]$ means the integer part of $x$ and we have defined
$z\equiv myy'/\hbar\tau$.
Now, using the definition of the Bessel functions
\beq
      J_n(z)\= \left({z\over2}\right)^n\,\sum_{p=0}^{+\infty}\,
       {(-1)^p\,z^{2p}\over 2^{2p}\, p!(n+p)!}~,
\eeq
we can show that
\beqn
    J_{n+1/2}(z)\= \sqrt{2\over\pi z}\,
      \Bigg\{\, \sum_{p=0}^{\lbrack {n\over2} \rbrack}\,
     {(-1)^p\,(n+2p)!\over (2p)!(n-2p)!(2z)^{2p}}\,{\rm sin}\left(
    z-{n\pi\over 2}\right)
    \nonumber\\
     + \sum_{p=0}^{\lbrack {n-1\over2} \rbrack}\,
     {(-1)^p\,(n+2p+1)!\over (2p+1)!(n-2p-1)!(2z)^{2p+1}}\,{\rm cos}\left(
    z-{n\pi\over 2}\right)\Bigg\}
\label{spherical}
\eeqn
Comparing equations \calle{prop1} and \calle{spherical} we derive the following
standard expression \cite{Grosche} for the propapagator of the potential
$V(y)={\hbar^2\over2m}{n(n+1)\over y^2}$:
\beq
\label{prop2}
     K(y,y';\tau)\={m\sqrt{yy'}\over\hbar \tau}\,i^{-(n+3/2)}\,
          e^{{im\over 2\hbar\tau}(y^2+y'^2)}\,
        J_{n+1/2}\left( {myy'\over\hbar\tau}\right)~.
\eeq
We can return to the original problem of constructing the propagator of
the radial oscillator potential \calle{radialoscillator} by making
use of the phase transformation \calle{phasetransformation} and
the substitution \calle{changevariables}. The final result is
\beq
     K(x,x';t)\={m\omega\sqrt{xx'}\over\hbar {\rm sin}\omega t
}\,i^{-(n+3/2)}\,
        {\rm exp}\left\{{im\omega\over2\hbar}{\rm cot}\omega t\,(x^2+x'^2)
                 \right\}
        J_{n+1/2}\left( {m\omega xx'\over\hbar{\rm sin}\omega t}\right)~.
\label{prop3}
\eeq
This is the standard form of the propagator found in literature \cite{Grosche}.

Taking into account the method of derivation presented above, it is
worthwhile to stress that time-dependent transformations may be very useful
even in the context of time-independent quantum mechanics.
This raises the question if all propagators of exactly soluble potentials
can be obtained by transforming the original problem to another simpler one
using time dependent transformations. Incidentally, notice that the harmonic
oscillator can be reduced to the problem of a free particle by using the
same transformation.

\acknowledgements
This work was supported in part by NSF.



\end{document}